\documentclass[11pt]{article}

\usepackage{amsmath,amsthm,amssymb}

\def\vertex{\circle*{2}}

\newtheorem{theorem}{Theorem}
\newtheorem{conjecture}{Conjecture}
\newtheorem{claim}{\ }[theorem]

\begin{document}
\title{On b-perfect chordal graphs\thanks{This research was supported
by Algerian-French program CMEP/Tassili 05 MDU 639.}}

\author{
Fr\'ed\'eric Maffray \thanks{%
CNRS, Laboratoire G-SCOP, 46 avenue F\'elix Viallet, 38031 Grenoble
Cedex, France.}
\and%
Meriem Mechebbek\thanks{USTHB, Laboratoire LAID3, BP32 El Alia, 
Bab Ezzouar 16111, Alger, Algeria.}%
}

\maketitle

\begin{abstract}
The b-chromatic number of a graph $G$ is the largest integer $k$ such
that $G$ has a coloring of the vertices in $k$ color classes such that
every color class contains a vertex that has a neighbour in all other
color classes.  We characterize the class of chordal graphs for which
the b-chromatic number is equal to the chromatic number for every
induced subgraph.
\end{abstract}

\section{Introduction} 

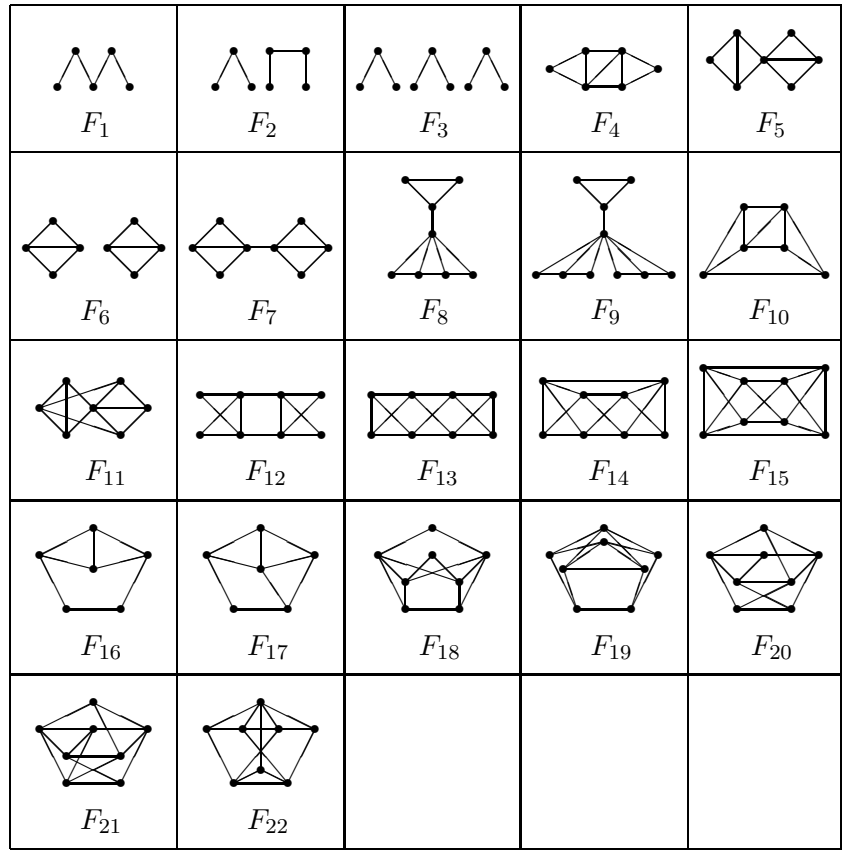
\begin{figure}[ht]
\unitlength=0.06cm
\begin{center}
\begin{tabular}{|c|c|c|c|c|}
     \hline
\begin{picture}(16,26) 
\multiput(0,12)(8,0){3}{\vertex}
\multiput(4,20)(8,0){2}{\vertex}
\multiput(0,12)(8,0){2}{\line(1,2){4}}
\multiput(8,12)(8,0){2}{\line(-1,2){4}}  
\put(5,2){$F_1$}
\end{picture}
&
\begin{picture}(20,26) 
\multiput(0,12)(8,0){2}{\vertex}
\multiput(12,12)(8,0){2}{\vertex}
\put(4,20){\vertex}
\multiput(12,20)(8,0){2}{\vertex}
\put(0,12){\line(1,2){4}} \put(8,12){\line(-1,2){4}}  
\multiput(12,12)(8,0){2}{\line(0,1){8}}
\put(12,20){\line(1,0){8}}
\put(7,2){$F_2$}
\end{picture}
&
\begin{picture}(32,26) 
\multiput(0,12)(12,0){3}{\vertex}
\multiput(8,12)(12,0){3}{\vertex}
\multiput(4,20)(12,0){3}{\vertex}
\multiput(0,12)(12,0){3}{\line(1,2){4}}
\multiput(8,12)(12,0){3}{\line(-1,2){4}}  
\put(13,2){$F_3$}
\end{picture}
&
\begin{picture}(24,26) 
\multiput(8,12)(8,0){2}{\vertex}
\multiput(8,20)(8,0){2}{\vertex}
\multiput(0,16)(24,0){2}{\vertex}
\multiput(8,12)(0,8){2}{\line(1,0){8}}
\multiput(8,12)(8,0){2}{\line(0,1){8}} 
\multiput(0,16)(16,4){2}{\line(2,-1){8}} 
\multiput(0,16)(16,-4){2}{\line(2,1){8}} 
\put(8,12){\line(1,1){8}}
\put(9,2){$F_4$}
\end{picture}
&
%
\begin{picture}(24,30) 
\multiput(6,12)(12,0){2}{\vertex}
\multiput(0,18)(12,0){3}{\vertex}
\multiput(6,24)(12,0){2}{\vertex}
\put(12,18){\line(1,0){12}}
\put(6,12){\line(0,1){12}}
\multiput(0,18)(6,6){2}{\line(1,-1){6}}
\multiput(12,18)(6,6){2}{\line(1,-1){6}}
\multiput(0,18)(6,-6){2}{\line(1,1){6}}
\multiput(12,18)(6,-6){2}{\line(1,1){6}}
\put(10,2){$F_5$}
\end{picture}
\\ \hline
\begin{picture}(30,30) 
\multiput(6,12)(18,0){2}{\vertex}
\multiput(0,18)(18,0){2}{\vertex}
\multiput(12,18)(18,0){2}{\vertex}
\multiput(6,24)(18,0){2}{\vertex}
\multiput(0,18)(18,0){2}{\line(1,0){12}}
\multiput(0,18)(18,0){2}{\line(1,-1){6}}
\multiput(6,24)(18,0){2}{\line(1,-1){6}}
\multiput(6,12)(18,0){2}{\line(1,1){6}}
\multiput(0,18)(18,0){2}{\line(1,1){6}}
\put(12,2){$F_6$}
\end{picture}
&
\begin{picture}(30,30) 
\multiput(6,12)(18,0){2}{\vertex}
\multiput(0,18)(18,0){2}{\vertex}
\multiput(12,18)(18,0){2}{\vertex}
\multiput(6,24)(18,0){2}{\vertex}
\put(0,18){\line(1,0){30}}
\multiput(0,18)(18,0){2}{\line(1,-1){6}}
\multiput(6,24)(18,0){2}{\line(1,-1){6}}
\multiput(6,12)(18,0){2}{\line(1,1){6}}
\multiput(0,18)(18,0){2}{\line(1,1){6}}
\put(12,2){$F_7$}
\end{picture}
&
\begin{picture}(18,39) 
\multiput(0,12)(6,0){4}{\vertex}
\multiput(9,21)(0,6){2}{\vertex}
\multiput(3,33)(12,0){2}{\vertex}
\put(0,12){\line(1,0){18}}
\put(3,33){\line(1,0){12}}
\put(9,21){\line(0,1){6}}
\put(0,12){\line(1,1){9}}\put(18,12){\line(-1,1){9}}
\put(6,12){\line(1,3){3}}\put(12,12){\line(-1,3){3}}
\put(9,27){\line(1,1){6}}
\put(9,27){\line(-1,1){6}}
\put(6,2){$F_8$}
\end{picture}
&
\begin{picture}(30,39) 
\multiput(0,12)(6,0){6}{\vertex}
\multiput(15,21)(0,6){2}{\vertex}
\multiput(9,33)(12,0){2}{\vertex}
\multiput(0,12)(18,0){2}{\line(1,0){12}}
\put(9,33){\line(1,0){12}}
\put(15,21){\line(0,1){6}}
\put(0,12){\line(5,3){15}}\put(30,12){\line(-5,3){15}}
\put(6,12){\line(1,1){9}}\put(24,12){\line(-1,1){9}}
\put(12,12){\line(1,3){3}}\put(18,12){\line(-1,3){3}}
\put(15,27){\line(1,1){6}}
\put(15,27){\line(-1,1){6}}
\put(12,2){$F_9$}
\end{picture}
& 
\begin{picture}(27,33) 
\multiput(0,12)(27,0){2}{\vertex}
\multiput(9,18)(9,0){2}{\vertex}
\multiput(9,27)(9,0){2}{\vertex}
\put(0,12){\line(1,0){27}}
\multiput(9,18)(0,9){2}{\line(1,0){9}}
\multiput(9,18)(9,0){2}{\line(0,1){9}}
\put(0,12){\line(3,2){9}} \put(27,12){\line(-3,2){9}}
\put(0,12){\line(3,5){9}} \put(27,12){\line(-3,5){9}}
\put(9,18){\line(1,1){9}}
\put(10,2){$F_{10}$}
\end{picture}
\\
\hline
\begin{picture}(24,30) 
\multiput(6,12)(12,0){2}{\vertex}
\multiput(0,18)(12,0){3}{\vertex}
\multiput(6,24)(12,0){2}{\vertex}
\put(12,18){\line(1,0){12}}
\put(6,12){\line(0,1){12}}
\multiput(0,18)(6,6){2}{\line(1,-1){6}}
\multiput(12,18)(6,6){2}{\line(1,-1){6}}
\multiput(0,18)(6,-6){2}{\line(1,1){6}}
\multiput(12,18)(6,-6){2}{\line(1,1){6}}
\put(0,18){\line(3,1){18}} \put(0,18){\line(3,-1){18}}
\put(10,2){$F_{11}$}
\end{picture}
&
\begin{picture}(27,27) 
\multiput(0,12)(9,0){4}{\vertex}
\multiput(0,21)(9,0){4}{\vertex}
\multiput(0,12)(0,9){2}{\line(1,0){27}}
\multiput(9,12)(9,0){2}{\line(0,1){9}}
\multiput(0,12)(18,0){2}{\line(1,1){9}}
\multiput(9,12)(18,0){2}{\line(-1,1){9}}
\put(10,2){$F_{12}$}
\end{picture}
&
\begin{picture}(27,27) 
\multiput(0,12)(9,0){4}{\vertex}
\multiput(0,21)(9,0){4}{\vertex}
\multiput(0,12)(0,9){2}{\line(1,0){27}}
\multiput(0,12)(27,0){2}{\line(0,1){9}}
\multiput(0,12)(9,0){3}{\line(1,1){9}}
\multiput(9,12)(9,0){3}{\line(-1,1){9}}
\put(10,2){$F_{13}$}
\end{picture}
&
\begin{picture}(27,30) 
\multiput(0,12)(9,0){4}{\vertex}
\multiput(9,21)(9,0){2}{\vertex}
\multiput(0,24)(27,0){2}{\vertex}
\multiput(0,12)(0,12){2}{\line(1,0){27}}
\put(9,21){\line(1,0){9}}
\multiput(0,12)(27,0){2}{\line(0,1){12}}
\multiput(0,12)(9,0){2}{\line(1,1){9}}
\multiput(18,12)(9,0){2}{\line(-1,1){9}}
\put(9,21){\line(-3,1){9}} \put(18,21){\line(3,1){9}}
\put(9,12){\line(-3,4){9}} \put(18,12){\line(3,4){9}}
\put(10,2){$F_{14}$}
\end{picture}
& 
\begin{picture}(27,33) 
\multiput(0,12)(27,0){2}{\vertex}
\multiput(9,15)(9,0){2}{\vertex}
\multiput(9,24)(9,0){2}{\vertex}
\multiput(0,27)(27,0){2}{\vertex}
\multiput(0,12)(0,15){2}{\line(1,0){27}}
\multiput(9,15)(0,9){2}{\line(1,0){9}}
\multiput(0,12)(27,0){2}{\line(0,1){15}}
\put(9,15){\line(1,1){9}} \put(18,15){\line(-1,1){9}}
\multiput(0,12)(18,3){2}{\line(3,4){9}}
\multiput(9,15)(18,-3){2}{\line(-3,4){9}}
\multiput(0,12)(18,12){2}{\line(3,1){9}}
\multiput(9,24)(18,-12){2}{\line(-3,1){9}}
\put(10,2){$F_{15}$}
\end{picture}
\\
\hline
\begin{picture}(24,36) 
\multiput(6,12)(12,0){2}{\vertex}
\multiput(0,24)(24,0){2}{\vertex}
\multiput(12,21)(0,9){2}{\vertex}
\put(6,12){\line(1,0){12}}
\put(12,21){\line(0,1){9}}
\put(18,12){\line(1,2){6}} \put(6,12){\line(-1,2){6}}
\put(12,21){\line(4,1){12}} \put(12,21){\line(-4,1){12}}
\put(12,30){\line(2,-1){12}} \put(12,30){\line(-2,-1){12}}
\put(9,2){$F_{16}$}
\end{picture}
&
\begin{picture}(24,36) 
\multiput(6,12)(12,0){2}{\vertex}
\multiput(0,24)(24,0){2}{\vertex}
\multiput(12,21)(0,9){2}{\vertex}
\put(6,12){\line(1,0){12}}
\put(12,21){\line(0,1){9}}
\put(18,12){\line(1,2){6}} \put(6,12){\line(-1,2){6}}
\put(12,21){\line(4,1){12}} \put(12,21){\line(-4,1){12}}
\put(12,30){\line(2,-1){12}} \put(12,30){\line(-2,-1){12}}
\put(12,21){\line(2,-3){6}}
\put(9,2){$F_{17}$}
\end{picture}
& 
\begin{picture}(24,36) 
\multiput(6,12)(12,0){2}{\vertex}
\multiput(6,18)(12,0){2}{\vertex}
\multiput(0,24)(12,0){3}{\vertex}
\put(12,30){\vertex}
\put(6,12){\line(1,0){12}}
\put(18,12){\line(1,2){6}} \put(6,12){\line(-1,2){6}}
\put(12,30){\line(2,-1){12}} \put(12,30){\line(-2,-1){12}}
\multiput(6,12)(12,0){2}{\line(0,1){6}}
\put(6,18){\line(3,1){18}} \put(18,18){\line(-3,1){18}}
\put(6,18){\line(1,1){6}} \put(18,18){\line(-1,1){6}}
\put(6,18){\line(-1,1){6}} \put(18,18){\line(1,1){6}}
\put(9,2){$F_{18}$}
\end{picture}
& 
\begin{picture}(24,36) 
\multiput(6,12)(12,0){2}{\vertex}
\multiput(3,21)(18,0){2}{\vertex}
\multiput(0,24)(24,0){2}{\vertex}
\multiput(12,27)(0,3){2}{\vertex}
\put(6,12){\line(1,0){12}}\put(3,21){\line(1,0){18}}
\put(6,12){\line(-1,3){3}} \put(18,12){\line(1,3){3}}
\put(6,12){\line(-1,2){6}} \put(18,12){\line(1,2){6}}
\put(3,21){\line(3,2){9}} \put(21,21){\line(-3,2){9}}
\put(3,21){\line(1,1){9}} \put(21,21){\line(-1,1){9}}
\put(0,24){\line(4,1){12}} \put(24,24){\line(-4,1){12}}
\put(0,24){\line(2,1){12}} \put(24,24){\line(-2,1){12}}
\put(9,2){$F_{19}$}
\end{picture}
& 
\begin{picture}(24,36) 
\multiput(6,12)(12,0){2}{\vertex}
\multiput(6,18)(12,0){2}{\vertex}
\multiput(0,24)(12,0){3}{\vertex}
\put(12,30){\vertex}
\multiput(6,12)(0,6){2}{\line(1,0){12}}
\put(18,12){\line(1,2){6}} \put(6,12){\line(-1,2){6}}
\multiput(6,18)(6,12){2}{\line(2,-1){12}} 
\multiput(6,12)(-6,12){2}{\line(2,1){12}}
\put(0,24){\line(1,0){24}}
\put(6,18){\line(-1,1){6}} \put(18,18){\line(1,1){6}}
\put(6,18){\line(1,1){6}}
\put(18,18){\line(-1,2){6}} 
\put(9,2){$F_{20}$}
\end{picture}
\\ \hline
\begin{picture}(24,36) 
\multiput(6,12)(12,0){2}{\vertex}
\multiput(6,18)(12,0){2}{\vertex}
\multiput(0,24)(12,0){3}{\vertex}
\put(12,30){\vertex}
\multiput(6,12)(0,6){2}{\line(1,0){12}}
\put(18,12){\line(1,2){6}} \put(6,12){\line(-1,2){6}}
\multiput(6,18)(6,12){2}{\line(2,-1){12}} 
\multiput(6,12)(-6,12){2}{\line(2,1){12}}
\put(0,24){\line(1,0){24}}
\put(6,18){\line(-1,1){6}} \put(18,18){\line(1,1){6}}
\put(6,18){\line(1,1){6}}
\put(18,18){\line(-1,2){6}} 
\put(6,12){\line(1,2){6}} 
\put(9,2){$F_{21}$}
\end{picture}
&
\begin{picture}(24,36) 
\multiput(6,12)(12,0){2}{\vertex}
\multiput(0,24)(8,0){4}{\vertex}
\multiput(12,15)(0,15){2}{\vertex}
\put(6,12){\line(1,0){12}}
\put(0,24){\line(1,0){24}}
\put(12,15){\line(0,1){15}}
\put(18,12){\line(1,2){6}} \put(6,12){\line(-1,2){6}}
\put(12,30){\line(2,-1){12}} \put(12,30){\line(-2,-1){12}}
\put(6,12){\line(5,6){10}} \put(18,12){\line(-5,6){10}}
\put(6,12){\line(2,1){6}} \put(18,12){\line(-2,1){6}} 
\put(8,24){\line(2,3){4}} \put(16,24){\line(-2,3){4}}
\put(9,2){$F_{22}$}
\end{picture}
& 
& 
& 
\\
\hline
\end{tabular}
\end{center}
\caption{Class ${\cal F}=\{F_1, \ldots, F_{22}\}$}
\label{fig1}
\end{figure}

We deal here with finite undirected graphs.  Given a graph $G$ and an
integer $k\ge 1$, a \emph{coloring} of $G$ with $k$ colors is a
mapping $c: V(G)\rightarrow \{1, \ldots, k\}$ such that any two
adjacent vertices $u,v$ in $G$ satisfy $c(u)\neq c(v)$.  For every
vertex $v$, the integer $c(v)$ is called the color of $v$.  The sets
$c^{-1}(1), \ldots, c^{-1}(k)$ that are not empty are called the color
classes of $c$.  A \emph{b-coloring} is a coloring such that every
color class contains a vertex that has a neighbour in each color class
other than its own, and we call any such vertex a \emph{b-vertex}.
The b-chromatic number $b(G)$ of a graph $G$ is the largest integer
$k$ such that $G$ admits a b-coloring with exactly $k$ colors.  The
concept of b-coloring was introduced in \cite{IM} and has been studied
among others in \cite{F,HK,KM,KZ,KTV}.  Let $\omega(G)$ be the maximum
size of a clique in a graph $G$, and let $\chi(G)$ be the chromatic
number of $G$.  It is easy to see that every coloring of $G$ with
$\chi(G)$ colors is a b-coloring, and so every graph satisfies
$\chi(G)\le b(G)$.  Ho\`ang and Kouider \cite{HK} call a graph $G$
\emph{b-perfect} if every induced subgraph $H$ of $G$ satisfies
$b(H)=\chi(H)$.  Also a graph $G$ is \emph{b-imperfect} if it is not
b-perfect, and \emph{minimally b-imperfect} if it is b-imperfect and
every proper subgraph of $G$ is b-perfect.  Ho\`ang, Linhares Sales
and Maffray \cite{HLM} found a list ${\cal{F}}$ of twenty-two
minimally b-imperfect graphs shown in Figure~\ref{fig1}, and posed the
following conjecture.
\begin{conjecture}[\cite{HLM}]
A graph is b-perfect if and only if it does not contain any member of
${\cal{F}}$ as an induced subgraph.
\end{conjecture}

Given a collection ${\cal{H}}$ of graphs, a graph $G$ is usually called
${\cal{H}}$-free if no induced subgraph of $G$ is a member of ${\cal{H}}$.
When ${\cal{H}}$ consists of only one graph $H$, we may write $H$-free
instead of $\{H\}$-free.  We let $P_k$ and $C_k$ respectively denote
the graph that consists of a path (resp.~cycle) on $k$ vertices.  We
use $+$ to denote the disjoint union of graphs, and $nF$ is the graph
which has $n$ components all isomorphic to $F$.  For example, $2K_2$
is the graph with two components of size $2$, and the first three
graphs in ${\cal{F}}$ are $P_5$, $P_4+P_3$ and $3P_3$.  We say that two
vertices $x,y$ in a graph $G$ are \emph{twins} if every vertex of
$G\setminus\{x, y\}$ that is adjacent to any of $x,y$ is adjacent to
both.  Note that two twins may be adjacent or not.

It is a routine matter to check that the graphs in class ${\cal{F}}$ are
b-imperfect and minimally so.  More precisely, for $i=1, 2, 3$, we
have $\chi(F_i)=2$ and $b(F_i)=3$, and $F_i$ admits a b-coloring with
$3$ colors in which its three vertices of degree $3$ have color $1, 2,
3$ respectively; and for $i=4, \ldots, 22$, we have $\chi(F_i)=3$ and
$b(F_i)=4$.

We will prove the conjecture in the case of chordal graphs.  Recall
that a graph $G$ is \emph{chordal} \cite{G,RR} if every cycle of
length at least four in $G$ has a chord (an edge between
non-consecutive vertices of the cycle).  We call \emph{hole} any
chordless cycle of length at least four.  In these terms, a graph is
chordal if and only if it is hole-free.
\begin{theorem}\label{thm:main}
Every ${\cal{F}}$-free chordal graph is b-perfect.
\end{theorem}
\noindent\emph{Proof of Theorem~\ref{thm:main}.} Suppose that the
theorem is false, and let $G$ be a counterexample to the theorem for
which $|V(G)|+|E(G)|$ is minimal.  Recall that, since $G$ is chordal,
it satisfies $\chi(G)=\omega(G)$ (see \cite{B,G}).  Since $G$ is a
counterexample to the theorem, it admits a b-coloring $c$ with $k\ge
\chi(G)+1 = \omega(G)+1$ colors.  For $i=1, \ldots, k$, let $u_i$ be
any b-vertex of color $i$, that is, a vertex that has a neighbour of
each color other than $i$.  Let $U=\{u_1, \ldots, u_k\}$.  Note that,
since $k>\omega(G)$, the set $U$ does not induce a clique.  As usual,
we say that a vertex is \emph{simplicial} if its neighbourhood induces
a clique.

\begin{claim}\label{uis}
For $i=1, \ldots, k$, vertex $u_i$ is not simplicial.
\end{claim}
\noindent\emph{Proof.} Suppose on the contrary and up to symmetry that
$u_1$ is simplicial.  Since $u_1$ is a b-vertex, it has a neighbour
$v_i$ of each color $i=2, \ldots, k$.  Then the set $\{u_1, v_2,
\ldots, v_k\}$ induces a clique of size $k>\omega(G)$, a
contradiction.  So Claim~\ref{uis} holds.

\begin{claim}\label{2K2}
$G$ contains a $2K_2$.
\end{claim}
\noindent\emph{Proof.} Suppose that $G$ contains no $2K_2$.  Since $U$
is not a clique, we may assume up to symmetry that $u_1, u_2$ are not
adjacent.  By Claim~\ref{uis}, vertex $u_1$ has two neighbours $v, v'$
that are not adjacent, and vertex $u_2$ has two neighbours $w, w'$
that are not adjacent.  Suppose that $u_1$ is adjacent to $w$.  Then
$u_1$ is not adjacent to $w'$, for otherwise $u_1, w, u_2, w'$ induce
a hole.  One of $v,v'$ is not equal to $w$, say $v\neq w$.  Also
$v\neq w'$ since $u_1$ is adjacent to $v$ and not to $w'$.  If $v$ is
not adjacent to $u_2$, then $v$ is adjacent to $w'$, for otherwise
$\{u_1, v, u_2, w'\}$ induces a $2K_2$; but then either $\{u_1, v, w',
u_2, w\}$ or $\{u_1, v, u_2, w\}$ induce a hole.  So $v$ is adjacent
to $u_2$.  Then $u_2$ is not adjacent to $v'$, for otherwise $\{u_1,
v, v', u_2\}$ induces a hole.  Then $v'$ is adjacent to $w'$, for
otherwise $\{u_1, v', u_2, w'\}$ induces a $2K_2$.  But then either
$\{u_1, v', u_2, w, w'\}$ (if $v', w$ are not adjacent) or $\{v', u_2,
w, w'\}$ (if $v', w$ are adjacent) induces a hole.  Therefore $u_1$ is
not adjacent to $w$.  Similarly, $u_1$ is not adjacent to $w'$, and
$u_2$ is not adjacent to any of $v, v'$.  Now $v$ must be adjacent to
$w$, for otherwise $\{u_1, v, u_2, w\}$ induces a $2K_2$, and by
symmetry, to $w'$ as well.  But then $\{v, u_2, w, w'\}$ induces a
hole, a contradiction.  So Claim~\ref{2K2} holds.

\

We say that a subgraph of $G$ is \emph{big} if it contains at least
two vertices.  Since $G$ contains a $2K_2$, it contains a set $S$ that
induces a subgraph with at least two big components and is maximal
with this property.  Let $R=V(G)\setminus S$.

\begin{claim}\label{xc1}
Every vertex of $R$ has a neighbour in every big component of $S$.
\end{claim}
\noindent\emph{Proof.} Suppose on the contrary that some vertex $x$ of
$R$ has no neighbour in some big component $C$ of $S$.  Then
$S\cup\{x\}$ induces a subgraph with at least two big components (of
which $C$ is one), which contradicts the maximality of $S$.  So
Claim~\ref{xc1} holds.

\begin{claim}\label{rk}
$R$ is a clique.
\end{claim}
\noindent\emph{Proof.} Suppose on the contrary that there are two
non-adjacent vertices $u,v$ in $R$.  Consider two big components $Z_1,
Z_2$ of $S$.  By Claim~\ref{xc1}, for each $i=1, 2$, $u$ has a
neighbour $u_i$ in $Z_i$ and $v$ has a neighbour $v_i$ in $Z_i$.
Since $Z_i$ is connected, we may choose $u_i, v_i$ and a path
$u_i$-$\cdots$-$v_i$ in $Z_i$ such that this path is as short as
possible (possibly $u_i=v_i$).  So no interior vertex of this path is
adjacent to $u$ or $v$.  But then the union of the two paths
$u_1$-$\cdots$-$v_1$, $u_2$-$\cdots$-$v_2$, plus $u$ and $v$, forms a
hole in $G$, a contradiction.  So Claim~\ref{rk} holds.

\begin{claim}\label{z}
There is a big component $Z$ of $S$ such that every vertex of $R$ is
adjacent to every vertex of every big component of $S\setminus Z$.
\end{claim}
\noindent\emph{Proof.} Suppose the contrary, that is, there are two
big components $Z_1, Z_2$ of $S$ and vertices $x_1, x_2$ of $R$ such
that $x_1$ has a non-neighbour in $Z_1$ and $x_2$ has a no-neighbour
in $Z_2$.  For each $i=1, 2$, since $Z_i$ is connected and by
Claim~\ref{xc1}, there are adjacent vertices $y_i, z_i$ in $Z_i$ such
that $x_i$ is adjacent to $y_i$ and not to $z_i$.  If $x_1=x_2$, then
$z_1$-$y_1$-$x_1$-$y_2$-$z_2$ is a $P_5$ in $G$, which contradicts
that $G$ is ${\cal{F}}$-free.  So $x_1\neq x_2$, and by the same argument
we may assume that $x_1$ is adjacent to all of $Z_2$ and that $x_2$ is
adjacent to all of $Z_1$.  By Claim~\ref{rk}, vertices $x_1, x_2$ are
adjacent.  Then $\{x_1, x_2, y_1, y_2, z_1, z_2\}$ induces an $F_4$,
which contradicts that $G$ is ${\cal{F}}$-free.  So Claim~\ref{z} holds.

\

Let $Z$ be a big component of $S$ as described in Claim~\ref{z}.  Let
$T=S\setminus Z$.  So $T$ contains a big component of $S$.  Put
$U_Z=U\cap Z$ and $U_T= U\cap T$.

\begin{claim}\label{rfar}
For every vertex $a\in R$ and every set $Y\subset Z$ that induces a
connected subgraph and contains no neighbour of $a$, there exists a
vertex of $Z$ that is adjacent to all of $Y\cup\{a\}$.
\end{claim}
\noindent\emph{Proof.} Pick any vertex $y$ in $Y$.  Since $Z$ is
connected, and $a$ has a neighbour in $Z$ by Claim~\ref{xc1}, there is
a shortest path $z_0$-$z_1$-$\cdots$-$z_p$ in $Z$ such that $z_0$ is
adjacent to $a$ and $z_p=y$.  Let $t$ be any vertex in a big component
of $T$.  By Claim~\ref{z}, vertices $a,t$ are adjacent.  Then $p=1$,
for otherwise $z_2$-$z_1$-$z_0$-$a$-$t$ is a $P_5$.  Thus $z_0$ is
adjacent to both $a, y$.  We show that $z_0$ is adjacent to all of
$Y$.  In the opposite case, since $Y$ is connected there are adjacent
vertices $y', y''$ such that $z_0$ is adjacent to $y'$ and not to
$y''$; but then $y''$-$y'$-$z_0$-$a$-$t$ is a $P_5$, a contradiction.
So Claim~\ref{rfar} holds.

\begin{claim}\label{rom}
$|R|\le \omega(G)-2$.
\end{claim}
\noindent\emph{Proof.} By the definition of $S$, the set $T$ contains
two adjacent vertices $a,b$.  By Claim~\ref{rk}, $R\cup\{a, b\}$ is a
clique.  So Claim~\ref{rom} holds.

\begin{claim}\label{zu}
$U_Z\neq \emptyset$.
\end{claim}
\noindent\emph{Proof.} Suppose on the contrary that $Z$ contains no
vertex of $U$.  Consider the graph $G' = G\setminus Z$.  Clearly, $G'$
is a chordal and ${\cal{F}}$-free graph, and $|V(G')| + |E(G')|< |V(G)|
+ |E(G)|$.  We show that $c$ is a b-coloring of $G'$.  To establish
this, consider vertex $u_i$ for each $i=1, \ldots, k$ and consider any
color $j\neq i$.  If $u_i$ is not in $R$, then $u_i$ has the same
neighbours in $G$ and in $G'$, so $u_i$ is a b-vertex in $G'$.  Now
suppose that $u_i$ is in $R$.  If $u_j$ is in a component of $S$ of
cardinality $1$, then $N(u_j)\subseteq R$, so $u_j$ is a simplicial
vertex by Claim~\ref{rk}, which contradicts Claim~\ref{uis}.  Thus
$u_j$ is in a big component of $T$.  Then $u_j$ is a neighbour of
$u_i$ by Claim~\ref{z} and the definition of $Z$.  Thus every $u_i$ is
a b-vertex for $c$ in $G'$.  But then $G'$ is a counterexample to the
theorem, which contradicts the minimality of $G$.  So Claim~\ref{zu}
holds.

\begin{claim}\label{sznop}
$T$ contains no $P_4$ and no $2P_3$.
\end{claim}
\noindent\emph{Proof.} Suppose on the contrary that $T$ contains a set
$Q$ of vertices that induces a $P_4$ or a $2P_3$.  Therefore $Z$
contains no $P_3$, for otherwise taking a $P_3$ in $Z$ plus $Q$ would
give an induced $F_2$ or $F_3$.  Since $Z$ is connected and contains
no $P_3$, it is a clique.  By Claim~\ref{zu}, we may assume that $u_1$
is in $Z$.  For $j=2, \ldots, k$, let $v_j$ be a neighbour of $u_1$ of
color $j$.  Since $\{u_1, v_2, \ldots, v_k\}$ is not a clique, we may
assume that $v_2, v_3$ are not adjacent.  Since $N(u_1)\subset R\cup
Z$ and both $R, Z$ are cliques, we may assume that $v_2\in R$ and
$v_3\in Z$.  By Claim~\ref{rom}, $R$ contains at most $k-3$ of the
$v_j$'s; so we may assume that $v_4\in Z$.  Now, if $v_2$ is not
adjacent to $v_4$, then $W\cup \{v_1, v_2, v_3, v_4\}$ induces an
$F_8$ or $F_9$; while if $v_2$ is adjacent to $v_4$ then the same set
contains an induced $F_5$.  So Claim~\ref{sznop} holds.

\begin{claim}\label{su}
$U_T\neq \emptyset$.
\end{claim}
\noindent\emph{Proof.} Suppose on the contrary that $T$ contains no
vertex of $U$.  Let $G'$ be the graph obtained from $G$ by removing
all edges whose two endvertices are in $T$.  Graph $G'$ satisfies
$|V(G')| + |E(G')|< |V(G)| + |E(G)|$ since we have removed at least
one edge because $T$ contains a big component of $S$.  We will show
that (a) $c$ is a b-coloring of $G'$, (b) $G'$ is a chordal graph, and
(c) $G'$ is ${\cal{F}}$-free.  These facts will imply that $G'$ is a
counterexample to the theorem, which will contradict the minimality of
$G$ and complete the proof of the claim.

To prove (a), it suffices to observe that every vertex of $U$ is a
b-vertex for $c$ in $G'$, because the edges we have removed from $G$
to obtain $G'$ are not incident with any vertex of $U$.  

To prove (b), observe that in $G'$ all vertices of $T$ are simplicial
(because their neighbourhood is in $R$) and thus cannot lie in a hole
of $G'$.  Moreover, $G'\setminus T = G\setminus T$.  So $G'$ contains
no hole and is chordal.

Now we prove (c).  Suppose on the contrary that $G'$ contains a member
$F$ of ${\cal{F}}$.  Note that $G'$ does not contain $F_i$ for $i=10,
\ldots, 22$, because every such $F_i$ contains a hole of length $4$ or
$5$, while $G'$ is chordal.  Thus $F$ must be one of $F_1, \ldots,
F_9$.  Graph $F$ must contain two vertices of $T$ that are adjacent in
$G$, for otherwise $F$ would be an induced subgraph of $G$.  Let $x,y$
be two vertices of $T$ in $F$ that are adjacent in $G$.  So $x,y$ lie
in the same big component of $T$, and it follows from Claim~\ref{z}
that the neighbourhood of each of them in $G'$ is $R$.  In particular,
in $F$ they are non-adjacent twins.  This immediately implies that $F$
cannot be $F_1$, $F_4$ or $F_8$ since such graphs do not have twins.
Thus $F$ must be one of $F_2$, $F_3$, $F_5$, $F_6$, $F_7$, $F_9$.
Note that, in each of these six cases, there is up to symmetry only
one pair of non-adjacent twins.

Suppose that $F$ is either $F_2$ or $F_3$.  So $F$ has vertices $x, y,
a, z_1, \ldots, z_p$, edges $xa, ya$, and either (if $F$ is $F_2$)
$p=4$ and $\{z_1, \ldots, z_4\}$ induces a $P_4$, or (if $F$ is $F_3$)
$p=6$ and $\{z_1, \ldots, z_6\}$ induces a $2P_3$ with edges $z_1z_2,
z_2z_3, z_4z_5, z_5z_6$.  As observed above, we may assume that
$x,y\in T$ and consequently $a\in R$; then vertices $z_1, \ldots, z_p$
are in a big component of $S$, and, by Claim~\ref{z}, they cannot be
in $T$, so they are in $Z$.  Let $p=4$.  By Claim~\ref{rfar}, $Z$
contains a vertex $z$ that is adjacent in $G$ to $a, z_1, \ldots,
z_4$.  Then $\{z, z_1, \ldots, z_4, a, x, y\}$ induces an $F_8$ in
$G$, a contradiction.  Now let $p=6$.  By Claim~\ref{rfar}, $Z$
contains a vertex $z$ that is adjacent in $G$ to $a, z_1, z_2, z_3$
and a vertex $z'$ that is adjacent in $G$ to $a, z_4, z_5, z_6$.  If
$z\neq z'$, then $\{z, z', z_1, \ldots, z_6\}$ induces an $F_6$ or
$F_7$ in $G$, a contradiction.  So $z=z'$.  But then $\{z, z_1,
\ldots, z_6, a, x, y\}$ induces an $F_9$ in $G$, a contradiction.

Suppose that $F$ is either $F_5$ or $F_9$.  So $F$ has vertices $x, y,
a, b, z_1, \ldots, z_p$, edges $xa, xb,$ $ya, yb,$ $ab,$ $az_1,$
$z_1z_2,$ $z_1z_3,$ $z_2z_3$ and either (if $F$ is $F_5$) $p=3$ and
$az_2$ is an edge, or (if $F$ is $F_9$) $p=6$ and vertices $z_4, z_5,
z_6$ induce a $P_3$ and are adjacent to $a$.  As observed above, we
may assume that $x,y \in T$ and consequently $a, b\in R$, and so $z_1,
\ldots, z_p\in Z$.  By Claim~\ref{rfar}, $Z$ contains a vertex $z$
that is adjacent in $G$ to $b, z_1, z_2, z_3$.  Then $z$ is adjacent
to $a$, for otherwise $\{z, a, b, z_1\}$ induces a hole in $G$.  But
then $\{z, a, b, z_1, z_3, x\}$ induces an $F_4$ in $G$, a
contradiction.

Finally suppose that $F$ is either $F_6$ or $F_7$.  So $F$ has
vertices $x, y,$ $a, b,$ $z_1, \ldots,$ $z_4$ and edges $xa, xb, ya,
yb, ab, z_1z_2, z_1z_3, z_1z_4, z_2z_3, z_2z_4$ and possibly (if $F$
is $F_7$) the edge $az_1$.  As observed above, we may assume that $x,y
\in T$ and consequently $a, b\in R$ and $z_1, \ldots, z_4\in Z$.  By
Claim~\ref{rfar}, $Z$ contains a vertex $z$ that is adjacent in $G$ to
$a, z_2, z_3, z_4$.  Vertex $z$ is also adjacent to $z_1$, for
otherwise $\{z, z_1, z_3, z_4\}$ induces a hole.  By Claim~\ref{rfar},
$Z$ contains a vertex $z'$ that is adjacent in $G$ to $b, z_1, \ldots,
z_4$.  If none of $z, z'$ is adjacent to both $a,b$, then either $\{z,
z', a, b\}$ or $\{z, z', a, b, z_2\}$ induces a hole.  So we may
assume, up to symmetry, that $z$ is adjacent to both $a,b$.  But then
$\{z, a, b, x, z_2, z_3, z_4\}$ induces an $F_5$ in $G$, a
contradiction.  Thus Claim~\ref{su} holds.

\begin{claim}\label{szk}
$U_T$ is a clique.
\end{claim}
\noindent\emph{Proof.} Suppose on the contrary that $u_1, u_2$ are non
adjacent vertices of $U_T$.  By Claim~\ref{uis}, vertex $u_1$ has two
neighbours $v, v'$ that are not adjacent, and vertex $u_2$ has two
neighbours $w, w'$ that are not adjacent.  By Claims~\ref{rk}
and~\ref{z} we have $v, v', w, w'\in T$.  If $u_1$ is adjacent to $w$,
then $\{u_1, w, u_2, w'\}$ induces a $P_4$ or a hole, which
contradicts Claim~\ref{sznop} or the chordality of $G$.  So $u_1$ is
not adjacent to $w$, and by symmetry it is not adjacent to $w'$, and
$u_2$ is not adjacent to any of $v, v'$.  If $v$ is adjacent to $w$,
then $\{v, u_1, v', w\}$ induce a $P_4$ or a hole, a contradiction.
So $v$ is not adjacent to $w$, and by symmetry it is not adjacent to
$w'$, and $v'$ is not adjacent to any of $w, w'$.  But now $\{u_1, v,
v', u_2, w, w'\}$ induces a $2P_3$, which contradicts
Claim~\ref{sznop}.  So Claim~\ref{szk} holds.

\

By Claim~\ref{su}, there is a vertex $u$ of $U$ in $T$.  By
Claim~\ref{uis}, vertex $u$ has two neighbours $t, t'$ that are not
adjacent.  By Claims~\ref{rk} and~\ref{z}, we have $t,t'\in T$.  In
other words, there is a $P_3$ $t$-$u$-$t'$ in $T$.

\begin{claim}\label{znop}
$Z$ contains no $P_4$ and no $2P_3$.
\end{claim}
\noindent\emph{Proof.} In the opposite case, a $P_4$ or $2P_3$ from
$Z$ plus the $P_3$ $t$-$u$-$t'$ from $T$ form an induced $F_2$ or
$F_3$ in $G$, a contradiction.  So Claim~\ref{znop} holds.

\begin{claim}\label{uzk}
$U_Z$ is a clique.
\end{claim}
\noindent\emph{Proof.} Suppose on the contrary that $u_1, u_2$ are non
adjacent vertices of $U_Z$.  Since $Z$ is connected, it contains a
path from $u_1$ to $u_2$, and since, by Claim~\ref{znop}, $Z$ contains
no $P_4$, such a path has length $2$, that is, $Z$ contains a vertex
$x$ adjacent to both $u_1, u_2$.  Suppose that some neighbour $y\neq
x$ of $u_1$ is not adjacent to $x$.  Then $y$ is also not adjacent to
$u_2$, for otherwise $\{y, u_1, x, u_2\}$ would induce a hole; and so
$y$-$u_1$-$x$-$u_2$ is a $P_4$.  If $y\in Z$ this contradicts
Claim~\ref{znop}, and if $y\in R$ then $u_2$-$x$-$u_1$-$y$-$t$ is a
$P_5$, another contradiction.  Therefore, $x$ is adjacent to every
neighbour of $u_1$ different from $x$, and similarly it is adjacent to
every neighbour of $u_2$ different from $x$.  By Claim~\ref{uis},
$u_1$ has neighbours $v, v'$ that are not adjacent.  Suppose that one
of $v,v'$, say $v$, is in $R$.  Then, since $R$ is a clique, $v'$ is
in $Z$, and, by the preceding argument, we have $x\neq v'$ and $x$ is
adjacent to $v, v'$.  But then $\{v, u_1, v', x, t, u, t'\}$ induces
an $F_5$, a contradiction.  Thus $v, v'$ are both in $Z$.  Likewise,
$u_2$ has neighbours $w, w'$ that are not adjacent, and they are both
in $Z$.  If $u_2$ is adjacent to $v$, then $u_2, v, u_1, v'$ induce
either a $P_4$ or a hole, a contradiction.  Thus $u_2$ is not adjacent
to $v$, and similarly not to $v'$, and $u_1$ is not adjacent to any of
$w, w'$.  Then $v$ is not adjacent to $w$, for otherwise
$u_1$-$v$-$w$-$u_2$ is a $P_4$.  Similarly, $v$ is not adjacent to
$w'$, and $v'$ is not adjacent to any of $w, w'$.  But now $\{u, t,
t', u_1, v, v', u_2, w, w'\}$ induces a $3P_3$ in $G$, a
contradiction.  So Claim~\ref{uzk} holds.

\

Let $C_T$ be the set of colors that appear in $U_T$.  By
Claim~\ref{su}, we have $|C_T|= |U_T|\ge 1$.  Let $C_Z$ be the set of
colors that do not appear in $R\cup U_T$.  By Claim~\ref{uis}, a
member of $U$ must be in a big component of $T$, and so, by
Claims~\ref{rk}, \ref{z} and~\ref{szk}, $R\cup U_T$ is a clique; thus
$|C_Z|\ge 1$.  Consider any color $j\in C_Z$.  By the definition of
$U$, every member of $U_T$ must have a neighbour of color $j$, and by
the definition of $C_Z$, any such neighbour must be in $T$.  Let $w_j$
be one vertex of color $j$ that is adjacent to the most members of
$U_T$.  So $w_j\in T$.  Suppose that $w_j$ has a non-neighbour $u'$ in
$U_T$.  Let $w'_j$ be a neighbour of $u'$ of color $j$.  So $w'_j\in
T$.  Since $u'$ is adjacent to $w'_j$ and not to $w_j$, the choice of
$w_j$ implies the existence of a vertex $u''$ of $U_T$ that is
adjacent to $w_j$ and not to $w'_j$.  But then $w_j$-$u''$-$u'$-$w'_j$
is a $P_4$, which contradicts Claim~\ref{sznop}.  Thus $w_j$ is
adjacent to all of $U_T$.  Now $R\cup U_T\cup \{w_j\}$ is a clique,
which implies $|C_Z|\ge 2$.  Let $W=\{w_j \mid j\in C_Z\}$.  Note that
$W$ is not a clique, for otherwise $R\cup U_T\cup W$ would be a clique
of size $k$ (because it contains a vertex of each color).

\ 

For each color $j\in C_Z$, the definition of $C_Z$ implies that $u_j$
is in $Z$.  So $$|U_Z|\ge |C_Z|\ge 2.$$ Consider any color $h\in C_T$.
By the definition of $U$, every member of $U_Z$ must have a neighbour
of color $h$, and by the definition of $C_T$ and by Claim~\ref{z}, any
such neighbour must be in $Z$.  Let $y_h$ be one vertex of color $h$
that is adjacent to the most members of $U_Z$.  So $y_h\in Z$.
Suppose that $y_h$ has a non-neighbour $u'$ in $U_Z$.  Let $y'_h$ be a
neighbour of $u'$ of color $h$.  So $y'_h\in Z$.  Since $u'$ is
adjacent to $y'_h$ and not to $y_h$, the choice of $y_h$ implies the
existence of a vertex $u''$ of $U_Z$ that is adjacent to $y_h$ and not
to $y'_h$.  But then $y_h$-$u''$-$u'$-$y'_h$ is a $P_4$, which
contradicts Claim~\ref{znop}.  Thus $y_h$ is adjacent to all of $U_Z$.
Let $Y=\{y_h \mid h \in C_T\}$.  So $|Y|= |C_T|$.  Suppose that $Y$ is
not a clique.  So there are non-adjacent vertices $y_g, y_h$ in $Y$.
Thus $|C_T|\ge 2$, and we have $u_g, u_h\in U_T$.  Recall that $W$ is
not a clique, so it contains two non-adjacent vertices $w_i, w_j$, and
by the definition of $W$ we have $u_i, u_j\in U_T$.  But then $\{y_g,
y_h, u_i, u_j, w_i, w_j, u_g, u_h\}$ induces an $F_6$, a
contradiction.  Thus $Y$ is a clique, and so $$\mbox{$Y\cup U_Z$ is a
clique of size at least $|C_T|+ |C_Z|\ge
3$.}$$

Let $R_1$ be the set of vertices of $R$ that have at most one
neighbour in $Y\cup U_Z$, and let $R_2=R\setminus R_1$.  If some
vertex $a\in R_2$ has a non-neighbour $v$ in $Y\cup U_Z$, then, since
$a$ has two neighbours $z, z'$ in $Y\cup U_Z$, we see that $\{a, z,
z', v, t, u, t'\}$ induces an $F_5$, a contradiction (recall that
$t$-$u$-$t'$ is a $P_3$ in $T$).  Thus every vertex of $R_2$ is
adjacent to every vertex of $Y\cup U_Z$.  This implies $R_1\neq
\emptyset$, for otherwise $R\cup Y\cup U_Z$ would be a clique of size
$k$ (because it contains a vertex of each color).

\

Consider any color $\ell$ that appears in $R_1$, and let $a_{\ell}$ be
the vertex of $R_1$ of color $\ell$.  By the definition of $U$ and
$R_1$, every vertex of $U_Z$, except possibly one, must have a
neighbour of color $\ell$ in $Z$.  Let $x_{\ell}$ be one vertex of $Z$
of color $\ell$ that is adjacent to the most members of $U_Z$.  By the
same argument as above concerning $y_h$, using the fact that $Z$
contains no $P_4$, we obtain that $x_{\ell}$ is adjacent to every
vertex of $U_Z$ that has a neighbour of color $\ell$ in $Z$.  Now we
show that $x_{\ell}$ is adjacent to all of $Y\cup U_Z$.  Suppose on
the contrary that $x_{\ell}$ has a non-neighbour $v$ in $Y\cup U_Z$.
If $x_{\ell}$ has two neighbours $z,z'$ in $Y\cup U_Z$, then either
$t$-$a_{\ell}$-$v$-$z$-$x_{\ell}$ is a $P_5$ (if $a_{\ell}$ is
adjacent to $v$), or $\{v, z, z', x_{\ell}, a_{\ell}, t, u, t'\}$
induces an $F_6$ or $F_7$, a contradiction.  So $x_{\ell}$ has only
one neighbour $z$ in $Y\cup U_Z$.  By the definition of $x_{\ell}$,
this implies that $U_Z=\{z, z'\}$ where $z'$ has no neighbour of color
$\ell$ in $T$.  Since $z'$ is in $U$, it must have a neighbour of
color $\ell$, and this can only be $a_{\ell}$.  But then
$x_{\ell}$-$z$-$z'$-$a_{\ell}$-$t$ is a $P_5$, a contradiction.  Thus
$x_{\ell}$ is adjacent to all of $Y\cup U_Z$.  Now we show that
$x_{\ell}$ is adjacent to all of $R_2$.  For suppose that $x_{\ell}$
is not adjacent to some vertex $a$ of $R_2$.  Let $z, z'$ be any two
vertices in $Y\cup U_Z$.  Then $\{x_{\ell}, z, z', a, t, u, t'\}$
induces an $F_5$, a contradiction.  In summary, $x_{\ell}$ is adjacent
to all of $Y\cup U_Z\cup R_2$.

\ 

Let $X=\{x_{\ell} \mid \mbox{color $\ell$ appear in $R_1$}\}$.  So
$X\neq \emptyset$.  Suppose that there are two non-adjacent vertices
$x_{\ell}, x_m$ in $X$.  Let $a_{\ell}$ be a vertex of color $\ell$ in
$R_1$.  Let $z, z'$ be any two vertices in $Y\cup U_Z$.  Then
$a_{\ell}$ is adjacent to $x_m$, for otherwise $\{x_{\ell}, x_m, z,
z', a_{\ell}, t, u, t'\}$ induces an $F_6$ or $F_7$.  Then $a_{\ell}$
is adjacent to $z'$, for otherwise
$x_{\ell}$-$z'$-$x_m$-$a_{\ell}$-$t$ is a $P_5$.  But then $\{x_m, z,
z', a_{\ell}, t, u, t'\}$ induces an $F_5$, a contradiction.
Therefore $X$ is a clique.  But now, $X\cup Y\cup U_Z\cup R_2$ is a
clique of size $k$ (because it contains a vertex of each color), a
contradiction.  This completes the proof of the theorem.  $\Box$

\

Theorem~\ref{thm:main} can be generalized slightly as follows.

\begin{theorem}\label{thm:c4}
Every ${\cal{F}}$-free $C_4$-free graph is b-perfect.
\end{theorem}
\noindent\emph{Proof.} Let $G$ be an ${\cal{F}}$-free $C_4$-free
graph.  Since $G$ contains no $P_5$, it contains no hole $C_k$ with
$k\ge 6$.  We prove that $b(G)=\chi(G)$ by induction on the number of
$C_5$'s contained in $G$.  If $G$ contains no $C_5$, then it is
chordal and the result follows from Theorem~\ref{thm:main}.  So we may
now assume that $G$ contains a $C_5$.  Let $z_1, \ldots, z_5$ be five
vertices such that, for $i=1, \ldots, 5$ modulo $5$, vertex $z_i$ is
adjacent to $z_{i+1}$ and not to $z_{i+2}$.  Let $Z=\{z_1, \ldots,
z_5\}$.  Let $x$ be a vertex of $G\setminus Z$ that has a neighbour in
$Z$.  If $x$ also has a non-neighbour in $Z$, then it is easy to see
that $Z\cup \{x\}$ contains a set that induces either a $P_5$, or a
$C_4$, or an $F_{16}$, a contradiction.  Thus $x$ is adjacent to all
of $Z$.  Let $X$ be the set of vertices that are adjacent to $Z$.
Note that $X$ is a clique, for if it contained two non-adjacent
vertices $x,y$, then $\{x, y, z_1, z_3\}$ would induce a $C_4$.
Suppose that $G$ admits a b-coloring $c$ with $k> \chi(G)$ colors.  We
may assume that the colors of $c$ that appear in $Z$ are $1, \ldots,
\ell$, with $3\le \ell\le 5$.  So only the colors $\ell+1, \ldots, k$
may appear in $X$.

If $\ell=3$, let $G'$ be the graph obtained from $G\setminus Z$ by
adding three new vertices $a_1, a_2, a_3$ that are pairwise adjacent
and all adjacent to all of $X$.  If $\ell=4$ or $5$, let $G'$ be the
graph obtained from $G\setminus Z$ by adding $\ell$ new vertices $a_1,
\ldots, a_{\ell}$ that are pairwise not adjacent and all adjacent to
all of $X$.  In either case, since $X$ is a clique the new vertices
$a_1, \ldots, a_l$ are simplicial, so they cannot belong to any hole,
and so $G'$ has strictly fewer $C_5$'s than $G$.

\begin{claim}\label{2b}
$b(G')\ge b(G)$.
\end{claim}
\noindent\emph{Proof.} Let $c'$ be the coloring of the vertices of
$G'$ defined by $c'(x)=c(x)$ if $x$ is a vertex of $G\setminus Z$ and
$c'(a_i)=i$ for $i=1, \ldots, \ell$.  Clearly, $c'$ is a coloring with
$k$ colors.  For each $i=1, \ldots, k$, let $u_i$ be a b-vertex of
color $i$ for $c$ in $G$.  Suppose that $u_i$ is in $G\setminus Z$.
Consider a neighbour $v_j$ of $u_i$ of color $j$ in $G$ for any $j\neq
i$.  Then either $v_j$ is in $G\setminus Z=G'\setminus Z$, and in this
case $v_j$ is a neighbour of $u_i$ of color $j$ in $G'$; or $v_j$ is
in $Z$, and in this case $j\in\{1, \ldots, \ell\}$ and $a_j$ is a
neighbour of $u_i$ of color $j$ in $G'$.  So $u_i$ is a b-vertex for
$G'$.  Now suppose that $u_i$ is in $Z$.  Then $u_i$ must have a
neighbour of every color $1, \ldots, \ell$ different from $i$, and
since such colors do not appear in $X$, they must appear in $Z$, and
so $\ell=3$ and all colors $4, \ldots, k$ appear in $X$.  Then $a_i$
is a b-vertex of color $i$ in $G'$.  Thus $c'$ has a b-vertex of every
color $i=1, \ldots, k$.  So Claim~\ref{2b} holds.

\begin{claim}\label{2g}
$\chi(G')\le \chi(G)$.
\end{claim}
\noindent\emph{Proof.} Consider any coloring $\gamma$ of $G$ with
$\chi(G)$.  We may assume that the colors of $\gamma$ that appear in
$Z$ are $1, \ldots, h$, with $3\le h\le 5$.  Let $\gamma'$ be defined
as follows.  For $x\in G\setminus Z$, set $\gamma'(x)=\gamma(x)$.  If
$\ell=3$, set $\gamma'(a_i)=i$ for $i=1, 2, 3$.  If $\ell=4$ or $5$,
set $\gamma'(a_i)=1$ for $i=1, \ldots, \ell$.  In either case,
$\gamma'$ is a coloring of $G'$ with at most $\chi(G)$ colors.  So
Claim~\ref{2g} holds.

\begin{claim}\label{2h}
$G'$ is ${\cal{F}}$-free and $C_4$-free.  
\end{claim}
\noindent\emph{Proof.} Suppose on the contrary that $G'$ contains a
subgraph $F$ which is either a member of ${\cal{F}}$ or a $C_4$.  Let
$A=\{a_1, \ldots, a_{\ell}\}$.  If $F$ contains at most two vertices
of $A$, then, since $Z$ has two adjacent vertices and also two
non-adjacent vertices, we can replace the vertices of $F\cap A$ by an
appropriate choice of vertices of $Z$ and we find a subgraph of $G$
that is isomorphic to $F$, a contradiction.  So $F$ must contain at
least three vertices of $A$.  Note that in $F$, the neighbourhood of
any of these vertices is equal to $F\cap X$, i.e., they are pairwise
twins.  But this is impossible, because no member of ${\cal{F}}\cup
\{C_4\}$ has three vertices that are pairwise twins.  Thus
Claim~\ref{2h} holds.

\

By Claims~\ref{2b}--\ref{2h}, $G'$ is an ${\cal{F}}$-free, $C_4$-free
graph with $b(G')\ge b(G)>\chi(G)\ge \chi(G')$ and $G'$ has strictly
fewer $C_5$'s than $G$, a contradiction.  This completes the proof of
Theorem~\ref{thm:c4}.  $\Box$


\end{document}